\documentclass[twocolumn,aps,prl,showpacs,amsmath,amssymb,psfrac,superscriptaddress]{revtex4}
\usepackage{amsmath}

\usepackage{epsfig}
\usepackage{graphicx}
\usepackage{dcolumn}
\usepackage{bm}

\renewcommand{\dag}{^{\dagger}}

\def\gl{\lower.35em\hbox{$\stackrel{\textstyle>}{\textstyle<}$}}
\def\gapp{\lower.35em\hbox{$\stackrel{\textstyle>}{\sim}$}}
\def\lapp{\lower.35em\hbox{$\stackrel{\textstyle<}{\sim}$}}

\begin{document}

\title{Localization-delocalization dichotomy: Inherent spectral properties of the cuprates}
\author{J.\ Ranninger}
\affiliation{Institut N\'eel, CNRS et Universit\'e Joseph Fourier, BP 166, 38042 Grenoble 
Cedex 09, France}
\author{T.\ Doma\'nski}
\affiliation{Institute of Physics, Marie Curie-Sk\l odowska University, 
20-031 Lublin, Poland}
\date{\today}

\begin{abstract} 
We consider hole pairing in the pseudopgap phase of High $T_c$ cuprates, as arising from 
resonant scattering on dynamically deformable molecular units. As a result, localized and 
delocalized features coexist in the one-particle spectra: the pseudogap and propagating 
diffusive Bogoliubov modes. Due to the anisotropy of the electron dispersion and pairing 
interaction, these two manifestations have different impact in the different regions of 
the Brillouin zone. We illustrate that for k-vectors crossing the arc, determined by the chemical potential, joining the anti-nodal and the nodal point.

\end{abstract}

\maketitle
{\it Introduction.} - We explore the spectral features of the  high $T_c$ cuprates within a 
scenario of resonant pairing, which characterizes systems close to a lattice fluctuation 
driven Superconductor - Insulator transition (SIT). The one-particle spectra are then determined  
by an interplay between localization and delocalization processes which originate on molecular 
scale [Cu-O-Cu] bonds \cite{Kohsaka-2007,Gomes-2007}. Together with their correlated deformable
ligand environments \cite{Lee-2006}, they act as potential pairing centers \cite{McElroy-2003} 
for the doped holes and break translational/rotational symmetry \cite{Kohsaka-2008} 
on a finite space-time scale. 
\begin{figure}[t]
  \begin{center}
    \includegraphics*[width=3in,angle=0]{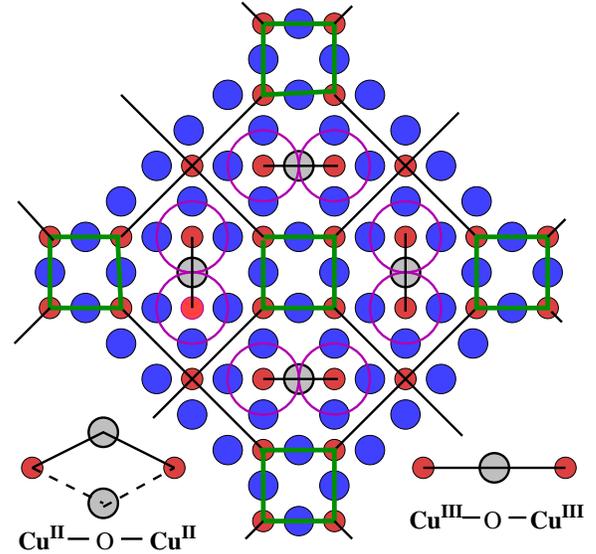}
    \caption{(Color online) Local structure of  the CuO$_2$ plane in form of (i)  Cu$_2$O$_7$ 
    domains acting as localizing pairing centers with directionally oriented Cu-O-Cu dumb-bells 
    with  central bridging O's (grey circles) and (ii) Cu$_4$ plaquettes (green square) housing 
    the delocalized charge carriers. Small red circles denote Cu cations and the larger blue 
    ones the O anions.}
\label{fig1}
\end{center}
\end{figure}
Systems, prone to dynamical lattice instabilities \cite{Ranninger-2008}, exist as  
metastable single phase solid solutions. Their kinetic stability is achieved  by 
synthesization at temperatures high 
enough such that the entropy contribution to the free energy stabilizes misfits between 
different stable cation-ligand complexes \cite{Sleight-1991}, which are then frozen in upon 
solidification. In the cuprates those misfits come from an incompatibility between Cu-O distances 
of stable square planar [Cu-O$_4$] configurations in the CuO$_2$ planes and of cation-ligand 
distances in the adjacent layers. 

Our contention is the following: The formal chemical Cu valence in the CuO$_2$ planes lies
between II and III (not to be confused with the ionic charge), which corresponds to stereochemical 
[Cu-O] distances in the [Cu-O$_4$] units of  1.94 $\AA$, respectively 1.84 $\AA$. The misfits between 
layers leads to dynamically unstable [Cu-O$_4$] units, linked by a bridging O which fluctuates in 
and out of the CuO$_2$ planes. This implies fluctuations between  kinked dumb-bells 
[Cu$^{II}$-O-Cu$^{II}$] (characteristic for  the undoped systems) and straight dumb-bells 
[Cu$^{III}$-O-Cu$^{III}$], which capture momentarily two holes in the doped cuprates. 
It results in local charge/deformation fluctuations which  break up the over-all homogeneous 
structure of the cuprates (Fig. 1 and scanning tunneling microscopy (STM) results, 
Figs. 4 and 5 in ref. 
\cite{Kohsaka-2008}) into: (i) checkerboard charge ordered directionally oriented Cu$_2$O$_7$ 
domains (three nearest neighbor Cu-Cu distances wide), which act as valence fluctuating driven 
pairing centers and (ii) a quadratic sublattice structure composed of  Cu$_4$ plaquettes on which 
the holes behave as delocalized strongly correlated entities,  subject to  
d$_{x^2 - y^2}$-wave pairing correlations \cite{Hirsch-1988,Altman-2002}. The local 
lattice fluctuations of the Cu$_2$O$_7$ domains prevent translational symmetry breaking, 
such that on a macroscopic level those materials exhibit an 
overall homogeneous crystal structure in a coarse grained sense \cite{Balatsky-2006}. 
Exceptionally (La$_{2-x}$Ba${x}$CuO$_4$ for x=1/8), local 
lattice deformations can lock together in a charge ordered phase, and in this way  impeach a  
superconducting state \cite{Valla-2006}). 

In the cuprates, resonant pairing driven by local dynamical lattice fluctuations finds its support 
in great a variety of experimental findings: the longitudinal optical (LO) Cu-O bond stretching 
mode of about 50 meV is strongly coupled to charge carriers near the {\it hotspot} anti-nodal 
points in the Brillouin zone (BZ) $[q_x,q_y]= [\pm \pi/2,0],[0, \pm \pi/2]$, where their 
pairing results in the pseudogap feature \cite{Lee-2006}. Upon entering the 
superconducting doping regime, coming from the insulating parent compound, this LO mode 
splits into two modes, separated by $\simeq$ 10 meV \cite{Reznik-2006}. It indicates a crystal 
lattice symmetry breaking linked to dynamical charge inhomogeneities. Pressure 
\cite{Haeflinger-2006}, isotope substitution studies \cite{Rubio-Temprano-2000} and  spatially 
resolved $d^2I/d V^2$-imaging \cite{Lee-2006} show concomitant (anticorrelated) modulations
of the pseudo-gap size and the frequency of this LO buckling mode. 

Resonant scattering of holes between the selftrapping Cu$_2$O$_7$ domains and the  Cu$_4$ plaquettes, 
making up a sublattice structure, induces pairing correlations on that latter. The holes 
acquire (i) localized features in the high energy sector of the one-particle spectrum and 
(ii)  delocalized low energy features of coherently propagating diamagnetic fluctuations. 
The first are manifest in the pseudogap phenomenon and  the latter in propagating 
strongly bound Cooper pairs. These features derive from competing amplitude and   
phase fluctuations of the order parameter with energy scales k$_B$T$^*$ and k$_B$T$_c$ and 
result in the anti-correlated $T_c - T^*$ relation, as the SIT is approached upon underdoping. 
Sometimes, this is interpreted as describing two different energy gaps \cite{Tesanovic-2008,Huefner-2008}. 

Resonant scattering  between itinerant fermionic charge carriers (holes or electrons) 
and bosonic tightly bound pairs of them near a SIT, 
can be accounted for by a phenomenological Boson Fermion Model (BFM).  This model  was 
originally proposed by one of us (JR) in the  early eighties in an attempt to describe the 
relatively abrupt cross-over between a weak coupling 
adiabatic electron-phonon mediated BCS superconductor and an insulating state of diffusive 
bipolarons in the strong coupling anti-adiabatic regime. The BFM describes itinerant fermions 
in chemical equilibrium with localized bound pairs (bipolarons) of them. It captures a 
situation of a single-component system, where at any given moment a certain percentage of the 
charge carriers is locally paired \cite{Ranninger-2008}. 
The superconducting, respectively insulating gap in such a system is centered 
at the chemical potential and determines the Fermi surface, which may be hidden. The opening 
of this gap does not depend on a particular set of Fermi wavevectors and hence is unrelated to 
any global translational symmetry breaking. The insulating state is  a Mott correlation 
driven phase of singlet-{\it bonding pairs} \cite{Cuoco-2006,Stauber-2007}. To what extent such
a phase fluctuation driven insulator could result in a Cooper-pair Wigner crystal, had been 
examined \cite{Tesanovic-2004,Pereg-Barnea-2006}. It predicted a  texturing coming from 
vortex-antivortex fluctuations rather than from the local lattice instabilities we evoke here. 

The anisotropy of pairing and of the hole dispersion in the CuO$_2$ planes marks the relative 
importance of localization versus delocalization of the charge carriers. This is clearly manifest, 
as we move along the arc in the BZ, which corresponds to the Fermi surface in the 
non-interacting system. Near the anti-nodal points strong pairing results from strong 
intra-{\it bonding pair} phase correlations between 
bound fermion pairs and their itinerant counterparts \cite{Domanski-1996,Domanski-2003b}. 
It leads to localization, which shows up in form of a pseudogap in the one-particle spectral 
properties and  destroys the Fermi surface. As one moves toward the nodal points, 
$[k_x,k_y] = [\pm \pi/2,\pm \pi/2]$, the intra-{\it bonding pair} phase correlations are 
weakened. It results in increased amplitude fluctuations and hence a reduction of the degree 
of localization, manifest in a reduced size of the pseudogap. At the same time, upon approaching 
the nodal points, the inter-{\it bonding pair} phase correlations are strengthened and with it
the spatial superconducting phase locking. This 
leads to Bogoliubov modes, which emerge out of localized phase uncorrelated 
singlet-{\it bonding} and {\it anti-bonding pairs}. We derive below these properties on the 
basis of the BFM, adapted to the specific anisotropic features of the cuprates. 


{\it The Model.} - The salient feature of the BFM is a charge exchange term which controls 
the transfer of electrons (holes) between real and momentum space \cite{Hanaguri-2008}.
\begin{equation}
H^{exch}_{BFM}=\frac{1}{\sqrt{N}}\sum_{{\bf k},{\bf q}}(g_{{\bf k},{\bf q}}b_{\bf q}\dag 
c_{{\bf q-k},\downarrow}c_{{\bf k},\uparrow}+H.c.).
\label{H_exch}
\end{equation}
It comprises localized bound electron (hole) singlet pairs ($b_{\bf q}^{(\dagger)}$) and 
their itinerant  counterparts, i.e., fermionic charge carriers ($c^{(\dagger)}_{{\bf k}\sigma}$), located respectively on the deformable Cu$_2$O$_7$ domains and the Cu$_4$ plaquettes.
We shall here consider effective sites, centered at the quadratic Cu$_4$ 
plaquettes and  which include the surrounding Cu$_2$O$_7$ domains. The effective BFM Hamiltonian 
for such a scenario then is: $H_{BFM} = H^0_{BFM} + H^{exch}_{BFM}$, where
\begin{equation}
H^0_{BFM}=\sum_{\bf k}(\varepsilon_{\bf k} -\mu)c_{\bf k}^{\dagger} c_{\bf k}+\sum_{\bf q} 
(E_{\bf q}-2\mu) b_{\bf q}^{\dagger} b_{\bf q}.
\label{H_0}
\end{equation}
The bare dispersion of the fermions and the intrinsically localized  bosons is $\varepsilon_k = \varepsilon^0_k$, respectively  $E_{\bf q}=2\Delta$.  The single component nature of such 
systems is enforced by a unique chemical potential $\mu$. We assume the standard anisotropic bare electron dispersion of the 
CuO$_2$ planes as $\varepsilon^0_{\bf k} = -2t[cos k_x + cos k_y] + 4t'cos k_x cos k_y$ 
with $t'/t=0.4$ and an anisotropic bare d-wave exchange coupling 
$g^0_{{\bf k},{\bf q}} = g[cos k_x  - cos k_y]\delta_{q,0}$. The interplay between the 
delocalizing  and the localizing effect is described by $H_{BFM}^{exch}$. It results in a 
competition  between local intra-pair correlations leading to an insulator and spatial 
inter-pair correlations leading to a superconductor, as the strength of the exchange coupling  
$g^0_{{\bf k},{\bf q}}$  decreases. Close to the SIT, the fermionic features have strong 
contributions coming from the bosonic particles and vice versa. Our aim therefore is to reformulate
this interacting Boson-Fermion mixture in terms of two effective commuting Hamiltonians, one describing 
purely fermionic excitations and one purely bosonic ones. The boson-fermion interaction thereby 
is absorbed into interdependent coupling constants by renormalizing $g_{{\bf k},{\bf q}}$ down to zero via a flow-equation renormalization procedure 
\cite{Wegner-1994,Glazek-1994}. For isotropic exchange coupling and fermion dispersions 
this problem has been studied previously \cite{Domanski-2001,Domanski-2003a,Stauber-2007}, predicting 
the pseudogap \cite{Ranninger-1995} and damped Bogoliubov modes \cite{Domanski-2003a} 
in angle resolved photoemission spectra. Both have since been verified experimentally \cite{expPG}.

{\it Localization versus delocalization.} - We now illustrate how the specific anisotropic pairing 
and dispersion in  the cuprate CuO$_2$ planes influence the spectral properties as one moves 
along the arc in the BZ, mentioned above. The flow 
equation technique projects at every step of this procedure the renormalized Hamiltonian onto 
the basic structure given by  $H^0_{BFM}$ in eq. \ref{H_0} plus 
renormalization induced fermion-fermion interactions $U_{\bf k,p}$ \cite{Domanski-2001}.
In this way, the various parameters $\varepsilon_{\bf k}(\ell), E_{\bf q}(\ell),  
g_{{\bf k},{\bf q}}(\ell), \mu(\ell)$, characterizing 
$H^0$ and $H_{exch}$ evolve as the flow parameter $\ell$  increases.  
The renormalization procedure starts with parameters corresponding to $\ell=0$, given 
by $\varepsilon_{\bf k}^0, 2\Delta$ and $g^0_{{\bf k},0}$. The chemical potential $\mu(\ell)$ is 
chosen at each step of the renormalization flow such as to fix a given total number of fermions 
and bosons. As a representative example we take $n_{tot} = \sum_{\sigma}n^F_{\sigma} + 2 n^B = 1$, 
such as to reproduce the appropriate size and shape of the CuO$_2$ plane Fermi surface. 
We concentrate on the spectral functions for electrons with wave vectors 
${\bf k}=|{\bf k}|[sin \phi,cos \phi]$, orthogonally intersecting the arc at various 
${\bf k}_F$, where the motion of the holes is essentially one dimensional. $\phi$ denotes the 
angle of those ${\bf k}$-vectors with respect to the line $[\pi,\pi]- [\pi,0]$ (see Fig.3).

The renormalization procedure consists of transforming the Hamiltonian in infinitesimal steps,
controlled by the differential equation $\partial_\ell H(\ell)=[\eta(\ell),H(\ell)]$. In its 
canonical form \cite{Wegner-1994}, $\eta(\ell)=[H_0(\ell),H(\ell)]$ is an anti-Hermitean generator. 
For details of the ensuing coupled non-linear differential equations for the various $\ell$ 
dependent parameters, we refer the reader to 
our previous work \cite{Domanski-2001,Domanski-2003a}. The flow of these parameters converges for  
$\ell \rightarrow \infty$ and results in two uncoupled  systems: one for the fermionic 
excitations and one for the bosonic ones with a fixed point Fermion dispersion 
$\varepsilon^*_{\bf k} = \varepsilon_{\bf k}(\ell \rightarrow \infty)$. The bare 
exchange coupling $g^0_{{\bf k},0}$, being equal to zero at the nodal point ($\phi = \pi/4$), 
steadily increases as one moves to the anti-nodal point ($\phi = 0$), where it achieves its 
maximal value, equal to g. As a consequence, $\varepsilon^*_{\bf k}$ remains 
essentially unrenormalized for ${\bf k}$ vectors crossing the arc near the nodal point. Upon 
approaching the anti-nodal point, on the contrary, it acquires an S-like inflexion at ${\bf k}_F$.
Upon reducing the temperature T, it becomes increasingly more pronounced 
and foreshadows the evolution into a true superconducting gap below $T_c$.

In order to obtain the renormalized spectral function for the charge carriers, a similar 
renormalization procedure is applied to the fermion and boson operators. Their effective 
structure is given by \cite{Domanski-2003a}
\begin{equation}
{c^{\dagger}_{-{\bf k},-\sigma}(\ell) \brack c_{{\bf k},\sigma}(\ell)} = u_{\bf k}(\ell) {c^{\dagger}_{-{\bf k},-\sigma} \brack c_{{\bf k},\sigma}} 
\mp \frac{1}{\sqrt N}\sum_{\bf q}v_{{\bf k},{\bf q}}(\ell){b^{\dagger}_{\bf q} c_{{\bf q+k}, \sigma} \brack  b_{\bf q} c^{\dagger}_{{\bf q-k}, -\sigma}} 
\end{equation}
The $\ell$ dependent parameters $u_{\bf k}(\ell), v_{\bf k}(\ell)$ are 
determined by corresponding flow equations and result in a fermionic spectral function of the form
\begin{eqnarray}
A({\bf k},\omega) = |u^*_{\bf k}|^{2} \delta \left( \omega\!+\!\mu\!-\!\varepsilon^*_{\bf k} 
\right) \qquad \nonumber \\
+ \frac{1}{N} \sum_{{\bf q}\neq{\bf 0}} \left( n_{\bf q}^{B} + n_{{\bf q}-{\bf k}\downarrow}^{F} 
\right) |v^*_{{\bf k},{\bf q}} |^{2} \delta ( \omega\!-\!\mu \!+\! \varepsilon^*_{{\bf q}
\!-\!{\bf k}} \!-\!E^*_{\bf q}), \label{spectral}
\end{eqnarray}
We illustrate in Fig.2 this spectral function for a characteristic region on the arc 
(corresponding to $\phi = 15^o$), separating the localization from the delocalization 
dominated regime for  g = 0.1 and $\Delta = 0.075$ in units of a nominal band width
$D = 8t$. For  $|{\bf k}| < |{\bf k|}_F$ it consists of essentially delocalized  
features (the first term) following the dispersion $\varepsilon^*_{\bf k}$. For  
$|{\bf k}|_F < |{\bf k}| < |{\bf k}|_0$, in a corresponding energy interval 
$[\varepsilon_F, \varepsilon_F  + \sqrt{\Delta^2 + Z g^2}]$, the delocalized features, given by 
$\varepsilon^*_{\bf k}$, are accompanied by localized ones (the second term). They 
reflect diffusively propagating images of localized bonding and anti-bonding states,
such as given by the Green's function in the atomic limit 
($t, t' = 0$) \cite{Domanski-1996,Domanski-2003b}, 
$G(i \omega_n)=1/[G^0(i \omega_n)^{-1} - \Sigma(i \omega_n)]$ with  
\begin{eqnarray}
\Sigma(i \omega_n) = {(1-Z) \; g^{2} \;(i \omega_n + \mu) \over 
[(i \omega_n + \mu)(i \omega_n - 2 \Delta + \mu) - Zg^2]},
\end{eqnarray}
\begin{figure}[t]
  \begin{center}
    \includegraphics*[width=3.5in,angle=0]{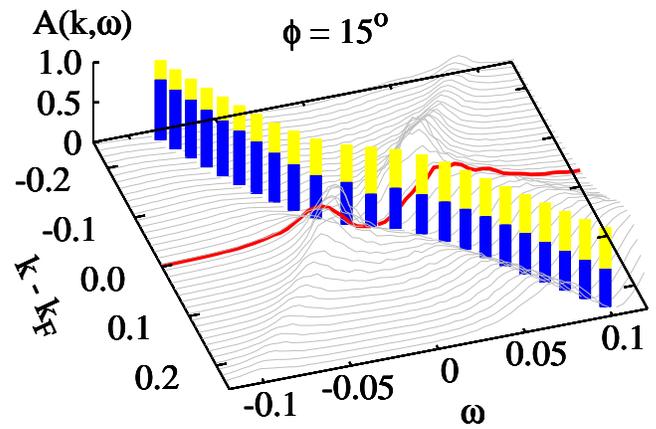}
    \caption{(Color online) $A({\bf k},\omega)$ at $T= 0.007$ ($< T^* = 0.016$) as a function 
    of $|{\bf k}|$ (in units of the inverse lattice vector) near ${\bf k}_F$ (red line), 
    corresponding to $\phi = 15^o$, orthogonally crossing the arc. The spectral weight of the 
    coherent and  incoherent contributions are indicated by blue, respectively yellow bars.}
\label{fig2}
\end{center}
\end{figure}
\begin{figure}[h]
  \begin{center}
    \includegraphics*[width=3.5in,angle=0]{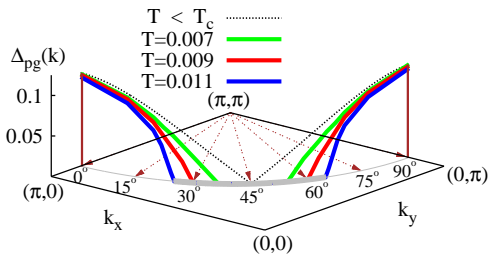}
    \caption{(Color online) Variation of the pseudogap for different k vectors, 
    orthogonally crossing the arc, given by angles $\phi$.} 
\label{fig3}
\end{center}
\end{figure}
having the structure of a BCS self-energy for localized Cooper pairs. 
$Z \simeq 2/[3+ \cosh (g/k_{B}T)]$ (for our choice of parameters) denotes the 
spectral weight of non-bonding delocalized charge carriers, described by 
$G^0(i \omega_n)= 1/(i \omega_n - \mu)$. In the normal state (constituted of an 
ensemble of spatially phase uncorrelated diffusively propagating {\it bonding} pairs) 
this atomic limit spectral feature dictates  (i) a broadened lower Bogoliubov band for 
$|{\bf k}| > |{\bf k}|_F$  and (ii) a broadened  upper Bogoliubov band for 
$|{\bf k}| < |{\bf k}|_F$, extending down to $k=0$, where it merges into the time 
reversed spectrum $-\varepsilon^*_{\bf k}$.  A  pseudogap gap in the partial density
of states, $\rho(\omega) = (1/N) \sum_{|{\bf k}|}A({\bf k}, \omega)$ opens up at some $T = T^*$ 
at ${\bf k}_F$. Its size, $\Delta_{pg}$, is determined by the distance between the peaks around 
$\varepsilon^*_{{\bf k}_F}$, when upon lowering T the deviation from the bare density of state, 
$\rho^0(\omega) = (1/N) \sum_{|{\bf k}|}\delta(\omega - \varepsilon_{\bf k})$ is 
reduced by $90 \%$. These features are attenuated, respectively enhanced, when  we move 
along the arc toward the nodal, respectively the antinodal, points. We illustrate 
in Fig. 3. the variation of $\Delta_{pg}$ for different T. Close to 
the anti-nodal point - the localization dominated regime - it  is relatively 
T independent. But approaching the nodal point it  abruptly drops to zero, already for 
finite values of $g^0_{{\bf k},0}$, which lets us envisage a BCS like onset of 
superconductivity (without any pseudogap) for  $60^o \geq \phi \geq 30^o$.  
Close to ${\bf k}_F$ the effect of  fermion-fermion interaction $U_{{\bf k},{\bf p}}$ is 
expected to change the poles of the coherent contributions into cut-singularities with 
complex residues, effectively showing up in  $A({\bf k}, \omega)$ as broad 
peaks \cite{Ranninger-1995,Ranninger-1996}. This effect of $U_{{\bf k},{\bf p}}$ could 
possibly result in a second gap \cite{Tesanovic-2008,Huefner-2008}, inside the rather T 
insensitive pseudogap, albeit that latter persisting into the superconducting phase upon 
lowering T. Concerning the evolution of the pseudogap phase into either a superconducting or
insulating one, we have to understand how these various parts of the BZ with dominating  
localizing, respectively delocalizing, effects are connected \cite{Perrali-2000}. This will depend on the conditions under which  the bosonic charge carriers condense into either a phase correlated or phase uncorrelated state. These questions will be addressed in future.

{\it Summary.}-The present scenario for the cuprates is based on resonant pairing, 
induced by local 
dynamical lattice instabilities. It makes use of the fact that such systems are prone to
segmentations which implies charge carriers existing simultaneously in (i) quasi localized 
states, originating from polaronic self-trapping in valence fluctuating local domains 
and (ii) delocalized states on a sublattice in which those domains are embedded. This scenario implies 
pairing correlations of varying strength in different regions of the BZ - a feature which
had been conjectured \cite{Anderson-1964} early on, trying to avoid the stringent limitations of  
$T_c$  within a BCS scenario.

\end{document}